\documentclass[12pt]{article}

\begin{document}
\topmargin 0pt
\oddsidemargin 0mm

\renewcommand{\thefootnote}{\fnsymbol{footnote}}
\begin{titlepage}
\begin{flushright}
IP/BBSR/2001-22\\
hep-th/0108174
\end{flushright}

\vspace{5mm}
\begin{center}
{\Large \bf Bound States of String Networks and D-branes}
\vspace{6mm}

{\large
Alok Kumar, Rashmi Rekha Nayak and\\
Kamal Lochan Panigrahi}\\
\vspace{5mm}
{\em Institute of Physics\\ 
Bhubaneswar 751 005\\ India\\
\vspace{3mm}

email: kumar, rashmi, kamal@iopb.res.in}

\end{center}
\vspace{5mm}
\centerline{{\bf{Abstract}}}
\vspace{5mm}
We show the existence of non-threshold bound states of  
$(p, q)$ string networks and $D3$-branes, preserving $1/4$ 
of the full type IIB supersymmetry, interpreted as string 
networks ``dissolved'' in $D3$-branes. 
We also explicitly write down the expression 
for the mass density of the system and discuss the extension 
of the construction to other $Dp$-branes. 
Differences in our construction of string networks with the ones
interpreted as dyons in $N=4$ gauge theories are also pointed out.

\end{titlepage}

\newpage
Non-threshold bound states of various D-branes 
\cite{witten,myers} 
have been objects of much interest due to 
their applications to the non-perturbative dynamics of string
theory and gauge theory, including from the point of view of AdS/CFT 
correspondence \cite{malda-russo,oz}. They have an
interpretation as branes that are ``dissolved'' inside other branes,
and preserve $1/ 2$ supersymmetry. 
They are also of importance in understanding 
the physics of black holes from a microscopic point of 
view \cite{maldacena}. Bound states of 
$F$-strings with $D$-branes have been analyzed as well 
\cite{russo-tsey,costa}. 
Such bound states are generally obtained by applying 
$T$-dualities \cite{pol} to delocalized brane solutions and have explicit
realizations as supergravity solutions. In view of their wide 
applications, it is of importance to analyze these results
further. 

In this note, we generalize the above constructions and obtain the 
bound states, now interpreted as 
$(p, q)$ string networks \cite{sen,sandip} ``dissolved'' 
in $D3$-branes. They preserve $1/4$ of the full type IIB supersymmetry and 
therefore describe new non-perturbative objects in these theories. 
It will also be pointed out later on that, our construction 
of the bound states of string networks and $D3$-branes 
are different from the ones appearing in the context of 
$N=4$ gauge theories, interpreted as dyons\cite{bergman,hata,hashi}. 

The existence of stable 
networks as well as web-like configurations for strings and branes is
now known for several years \cite{schwarz2} on the
basis of charge conservation, tension balance and supersymmetry analysis.
Although for large number of these configurations no explicit supergravity
or worldvolume realizations are  known,  
several examples in the context of string networks have been 
worked out from world volume point of view\cite{dasg,hata}.
Results in our paper give evidence for the existence of similar
configurations when they are dissolved inside other D-branes.

We now start by writing down the classical 
supergravity solution \cite{malda-russo}
corresponding to the $D1-D3$ bound state\cite{myers,costa}, 
preserving ${1/2}$ supersymmetry :
\begin{eqnarray}
&&ds^2_{str} = f^{-1/2} [ - dx_0^2 + dx_1^2 +{h}(dx_2^2 + dx_3^2 )]
+ f^{1/2} ( dr^2 + r^2 d \Omega^2_5 ), \nonumber\\
&&f = 1 + {{\alpha'}^2 R^4\over r^4} ,\
~~~h^{-1} =\sin^2\phi f^{-1}+\cos^2\phi, \nonumber\\
&&B_{23} ={\sin\phi\over \cos\phi }\ \ f^{-1} h, 
~~~e^{2\Phi} = g^2 h, \nonumber\\
&&F_{01r} = {1 \over g} \sin\phi\ \partial_r f^{-1}, \nonumber\\
\ \ \ \ \
&&F_{0123r} = {1 \over g} \cos\phi \ h \ \partial_r f^{-1},
\label{d1-d3-bound}
\end{eqnarray}
where $B_{\mu \nu}$ is the NS-NS antisymmetric tensor field. 
$F_{\mu \nu \rho}$ and $F_{\mu \nu \rho \alpha \beta}$ are 
respectively R-R 3-form and 5-form field strengths. 
The asymptotic value of the B field in eqn. (\ref{d1-d3-bound})
is $B_{23}^{\infty} = tan(\phi)$ and gives the 
expression for the ratio of charge densities of 
(smeared) $D1$ and $D3$ branes. 
The parameter R is defined by $cos \phi R^4 = 4\pi g n$,
with $n$ being the number of $D3$-branes. Finally $g \equiv g_{\infty}$ is
the asymptotic value of the string coupling.

To describe explicitly  the ${1/2}$ supersymmetry property of the 
$D1-D3$ bound state, we note from their explicit solution in 
eqn.(\ref{d1-d3-bound}) that, they also have an alternative
interpretation in terms of $D3$-branes in a constant NS-NS antisymmetric 
tensor background of magnetic type:
$B_{23}= tan(\phi)$. The ${1/2}$ supersymmetry 
condition is then written in the following form \cite{hashi}:
\begin{equation}
(\epsilon_L - \epsilon_R) = sin\phi ~\Gamma^{0 1} 
              (\epsilon_L - \epsilon_R) + 
           cos\phi ~\Gamma^{0 1 2 3} (\epsilon_L + \epsilon_R), 
\end{equation}
where $\epsilon_L$ and $\epsilon_R$ are two positive chirality 
space-time spinors arising from the left and the right moving sectors
of the type IIB string theory. The above condition can also be 
written in an alternative form:
\begin{equation}
\epsilon_L = - sin\phi ~\Gamma^{0 1} \epsilon_R 
              + cos \phi ~\Gamma^{0 1 2 3} \epsilon_R, \\
\label{d1-d3}
\end{equation}
or equivalently,
\begin{equation} 
\epsilon_R = - sin\phi ~\Gamma^{0 1} \epsilon_L 
              - cos \phi ~\Gamma^{0 1 2 3} \epsilon_L.
\label{d1-d3-2}
\end{equation}

Supersymmetry conditions of eqns. (\ref{d1-d3}) and 
(\ref{d1-d3-2}) reduce to that of a standard $D3$-brane
for $\phi= 0$. From eqn. (\ref{d1-d3-bound}), we also notice that 
$D$-string in the above bound state lies along the $x^1$ axis
and is smeared in the remaining spatial 
directions $x^2$ and $x^3$, giving it 
the interpretation of a $D$-string being ``dissolved'' in a 
$D3$-brane. A generalization of the supergravity solution in
eq. (\ref{d1-d3-bound}), representing $((F, D1),D3)$ bound state,
is known and corresponds to the case when both electric and 
magnetic type $B_{\mu \nu}$ fields are turned on 
\cite{malda-russo,jabbari}. These solutions
also preserve $1/2$ supersymmetry, thereby ensuring their
stability. 

The mass-density of these $((F1, D1), D3)$ bound states can be 
expressed (in string-frame) as \cite{myers,lu2,cai}:
\begin{equation}
m^2 = T_0^2\left[{n^2\over g^2} + |p + q \tau|^2 \right], 
\label{energy-density}
\end{equation}
where we have one $(p, q)$-string along, say $x^1$ direction
per $(2\pi)^2 \alpha'$ area \cite{lu1} over the $x^2-x^3$ plane,
and $n$ is the  $D3$-brane charge. 
Also, $T_0 = {1\over (2 \pi)^3 \alpha'^2}$ and 
axion-dilaton moduli are given as: 
$\tau \equiv \chi + {i \over g}$. We also notice that mass-density
(\ref {energy-density}), is a sum of distinct energy densities,
associated with a $(p, q)$-string and that of a $D3$-brane. Moreover
the contributions of $(p, q)$-string for different $(p, q)$'s
remain identical to the one, when $D3$-brane is absent. 

We now discuss the construction of the bound state of 
string networks and $D3$-brane from supersymmetry point of view.
Following above reasoning, these objects can also be viewed 
as $(p, q)$ string networks dissolved in a $D3$-brane.
To discuss the network construction we now complexify eqns.(\ref
{d1-d3}) and (\ref{d1-d3-2}):  
\begin{equation}
(\epsilon_L - i \epsilon_R) = i sin\phi  ~\Gamma^{0 1} 
              (\epsilon_L + i \epsilon_R) + 
           i cos\phi  ~\Gamma^{0 1 2 3} (\epsilon_L - i \epsilon_R),
\label{complex-d1-d3}
\end{equation}
giving the $1/2$ supersymmetry projection of a 
$(D1- D3)$-bound state. Then, 
to write down the supersymmetry projection of a bound state 
$((F1, D1), D3)$ of a  
$(p, q)$-string and $D3$-brane, we use the fact that they can be 
generated by applying $SL(2, Z)$ duality \cite {schwarz} 
on the $D1-D3$ bound state discussed above. This procedure also gives the 
$1/2$ supersymmetry condition for the $((F1, D1), D3)$
bound state, by using that
the spinors, $(\epsilon_L \pm i \epsilon_R)$, transform covariantly 
under the maximal compact subgroup, $SO(2) \in SL(2, R)$, with 
$SL(2, R)/SO(2)$ parametrizing the moduli space represented by 
axion-dilaton fields. The transformation properties of spinors 
are given as:
\begin{equation}
(\epsilon_L \pm i \epsilon_R) \rightarrow e^{i {\alpha\over 2}} 
(\epsilon_L \pm i \epsilon_R).
\end{equation}
To obtain the phase $\alpha$ for a given $SL(2, Z)$ transformation, 
one notes that using the vielbein $E$, corresponding to the 
axion-dilaton moduli ${\cal M} \equiv E E^T$, any $SL(2, R)$ vector can be 
turned into an $SO(2)$ vector. As a result, the phase transformation 
parameter $\alpha$ can be read off from the corresponding $SL(2, Z)$ 
parameters. The supersymmetry condition for a 
$(p, q)$-string dissolved in a
$D3$ brane can then be generated from the one 
in (\ref{complex-d1-d3}) and has a form:
\begin{equation}
(\epsilon_L - i \epsilon_R) = e^{i \Theta{(p, q, \tau)}} 
             sin\phi  ~\Gamma^{0 1} 
              (\epsilon_L + i \epsilon_R) + 
           i cos\phi  ~\Gamma^{0 1 2 3} (\epsilon_L - i \epsilon_R).
\label{complex-p-q}
\end{equation}
We notice that a phase factor $\Theta$, dependent on axion- dilaton
moduli $(\tau)$ as well as $SL(2, Z)$ quantum numbers $(p, q)$:
$e^{i \Theta (p, q, \tau)} = {{p + q \tau}/|{p + q \tau}|}$,   
appears in the first term in the R.H.S. representing the supersymmetry 
condition of a $(p, q)$ string. Since $D3$-branes are $SL(2, Z)$
invariant objects, the second term in the R.H.S. 
of eqn. (\ref{complex-p-q}) remains unchanged
with respect to the one for $D1 - D3$ case. 

Now, to show the possibility of a string network construction, 
we consider a $(p, q)$ string lying in $x^1 - x^2$ plane at an angle 
$\theta$ with the $x^1$ axis. Then equation (\ref{complex-p-q}) is 
replaced by:
\begin{equation}
(\epsilon_L - i \epsilon_R) = e^{i \Theta{(p, q, \tau)}} 
             sin\phi ~\Gamma^{0}(\Gamma^1 ~cos\theta + \Gamma^2 ~sin\theta) 
              (\epsilon_L + i \epsilon_R) + 
           i cos\phi ~\Gamma^{0 1 2 3} (\epsilon_L - i \epsilon_R).
\label{complex-p-q-1}
\end{equation}

It can be seen that, as in the case of string networks in the absence
of $D3$-brane,  
if one identifies the orientation of the $(p, q)$-
string inside $D3$-brane, with its phase in the internal space:
$\theta = \Theta(p, q, \tau)$, then the above supersymmetry 
condition is solved by the following projections:
\begin{equation}
(\epsilon_L - i \epsilon_R) =  sin\phi ~\Gamma^{0 1} 
              (\epsilon_L + i \epsilon_R) + 
           i cos\phi ~\Gamma^{0 1 2 3} (\epsilon_L - i \epsilon_R),
\label{F-string}
\end{equation}
and
\begin{equation}
(\epsilon_L - i \epsilon_R) = i sin\phi ~\Gamma^{0 2} 
              (\epsilon_L + i \epsilon_R) + 
           i cos\phi ~\Gamma^{0 1 2 3} (\epsilon_L - i \epsilon_R).
\label{D-string}
\end{equation}
We notice that the projection condition (\ref{F-string}) corresponds
to that of an $F$-sting along $x^1$ axis dissolved in a $D3$-brane. 
Similarly, the projection condition
(\ref{D-string}) corresponds to that of a $D$-string along $x^2$ 
axis, dissolved in the same $D3$-brane. These together imply that
supersymmetry is broken to $1/4$ of the original one. Interestingly, 
the supersymmetry condition, eqn.(\ref{complex-p-q-1}), 
is satisfied for arbitrary $(p, q)$ with 
only finite number of projections, provided the above identification
of the phases, $\theta = \Theta(p, q, \tau)$ holds. 
The projection conditions, eqns. (\ref{F-string}) 
and (\ref{D-string}), also reduce to the ones in \cite{sen}, for 
$\phi = {\pi \over 2}$, which corresponds to the case when 
there is no $D3$-brane.  
We have therefore shown the existence of a 
$((p, q) string~network, D3)$,  bound state preserving $1/4$ 
supersymmetry. 

To confirm the $1/4$ supersymmetry property of our configuration 
further, we now show that 
simultaneous solutions for $\epsilon_L$ and $\epsilon_R$, 
of appropriate type, do exist for eqns. (\ref{F-string}) 
and (\ref{D-string}). In this connection, we note that
eqn. (\ref{F-string}), representing 
the supersymmetry of $F$-string dissolved in $D3$-
brane, can be written as 
\begin{equation}
\epsilon_L = sin \phi~\Gamma^{0 1} \epsilon_L 
            + cos \phi~\Gamma^{0 1 2 3}\epsilon_R,
\label{d3-F}
\end{equation}
or equivalently as
\begin{equation}
\epsilon_R = - sin \phi~\Gamma^{0 1} \epsilon_R 
             - cos \phi~\Gamma^{0 1 2 3}\epsilon_L.
\label{d3-F-2}
\end{equation}
The supersymmetry conditions of dissolved $D$-strings, along $x^1$,
were already written in eqn. (\ref{d1-d3}), or (equivalently in) 
(\ref{d1-d3-2}).
From eqn. (\ref{D-string}), we get conditions that are 
identical to the ones in eqns. (\ref{d1-d3}) and (\ref{d1-d3-2}),
when we replace $\Gamma^{0 1}$ by $\Gamma^{0 2}$.
In particular, for our argument we use:
\begin{equation} 
\epsilon_R = - sin\phi ~\Gamma^{0 2} \epsilon_L 
              - cos \phi ~\Gamma^{0 1 2 3} \epsilon_L,
\label{d1-d3-3}
\end{equation}
as well as eqn. (\ref{d3-F}) as independent conditions
following from (\ref{F-string}) and (\ref{D-string}). Now, substituting 
$\epsilon_R$ from eqn. (\ref{d1-d3-3}) into
(\ref{d3-F}), one gets:
\begin{equation}
\epsilon_L = \Gamma^0(\Gamma^1~sin\phi - \Gamma^3~cos\phi)\epsilon_L.
\label{F-D-string}
\end{equation}
The $1/4$ supersymmetry now directly follows from eqns.(\ref{d1-d3-3})
and (\ref{F-D-string}). 

We now obtain the mass density of 
the $(string~network, D3)$ bound state 
that we have constructed. For this purpose, one can start with the
expression of the mass density for a bound state of 
$(p, q)$-string with $D3$ branes as in eqn.(\ref{energy-density}).
As already emphasized, $(p, q)$-strings inside $D3$-branes have
distinct contribution to the total mass formula. 
Now, for the case of $(string~network, D3)$ bound state
the contribution to the total mass, coming from the network, 
can be written as sum of
contributions from different strings in that network\cite{sen}: 
\begin{equation}
m_{network}^2 = (\Sigma_i l_i T_i)^2,
\end{equation}
where $l_i$'s are the the lenghts of various links and $T_i$'s are 
the corresponding tensions. Final expression for mass density
is then obtained by
adding contribution from the $D3$-branes as well. In other words, 
the modification to the mass formula in the string network case
is essentially due to the replacement of the $(p, q)$-string tension, by 
the corresponding network mass formula in eqn. (\ref{energy-density}).

To write the expression for the mass density in a concrete form, 
we consider the case when the string network as well as 
$D3$-brane are wrapped on a $T^2$.
In this context, for the wrapping of the string 
network, one defines lattice vectors $\vec{a}, \vec{b}$, 
constructed out of the link vectors $\vec{l_i}$,$(i = 1, 2, 3)$ 
of a $3$-prong string junction in a periodic string network 
of strings with quantum numbers $(p_i, q_i)$, $(i=1,2,3)$
\cite{sen,kumar}, obeying charge conservation on the junction. 
The $T^2$ is parametrized by moduli:
$\lambda_1 = \vec{a}.\vec{b}/\vec{a}^2$, 
$\lambda_2 = |\vec{a}\times \vec{b}|/\vec{a}^2 $.
Total mass density (per unit length)
then turns out to be of the form (now in Einstein-frame):
\begin{equation}
m^2 = T_0^2 n^2 A^2 + m_{network}^2, 
\label{total-energy}
\end{equation}
where $A = |\vec{a}\times \vec{b}|$ is the area of $T^2$.
Explicit expression for network contribution to this mass-density:
$m_{network}^2$ is identical to the one 
in \cite{sen}, with appropriate replacements of charges by 
charge-densities, coming from the smearing
of the resulting (delocalized) particle-like state in the 
unwrapped direction of the $D3$ brane. Putting the factors of 
$\alpha'$ etc. appropriately:
\begin{equation}
m_{network}^2  = {1\over (2 \pi)^4 \alpha'^3} A 
             \pmatrix{p_1 & q_1 & p_2 & q_2}
 (M \pm L) \pmatrix{p_1
\cr q_1\cr p_2 \cr q_2},\
\label{mass-sq}
\end{equation} 
where we have one wrapped string network per $2\pi \sqrt{\alpha'}$ length 
along the unwrapped direction of $D3$-brane. Also, 
\begin{eqnarray}
M = {1\over \lambda_2}\pmatrix{ \cal M & \lambda_1 \cal M \cr
\lambda_1 \cal M & |\lambda|^2 \cal M}\, , 
~~~~~L = \pmatrix{0 & \cal L \cr -\cal L & 0},\\
{\cal M} = {1\over \tau_2}\pmatrix{1 & \tau_1 \cr \tau_1 & |\tau|^2}\, ,
~~~~~\cal L = \pmatrix{ 0 & 1 \cr -1 & 0}\, .
\end{eqnarray}

We have therefore given the expression for the mass density of the 
bound states discussed above. We also notice that the above 
mass formula reduces to the one for 
the conventional string networks \cite{sen} 
in the absence of $D3$-branes, $(n=0)$. Moreover,
by setting charges $(p_2, q_2) = (0, 0)$, which implies the reduction to 
the case of a straight string with $(p_1, q_1)$ charges, one can obtain the 
energy spectrum of $((F, D1), D3)$ bound state with $1/2$ supersymmetry.  
In the compactified theory, the mass formula (\ref{total-energy})
also corresponds to that of a bound state of a particle with $U$-duality 
charges, dissolved in a string with, $(0, 0, 1)$ charge. 
Our result therefore turns out to be consistent with a 
general supersymmetry analysis of BPS objects in eight dimensions 
\cite{malda-ferra} implying $1/4$ supersymmetry for
particle-like objects and $1/2$ for string-like ones in $D = 8$.  
Apart from the supersymmetry analysis that we have presented in
the paper, we also
note that the non-threshold BPS mass formula, written in 
eqn. (\ref{total-energy}), implies that such bound states 
of string-networks and D3-branes are also 
energetically favorable, and therefore likely to be formed, 
when several $(p, q)$-strings are dissolved 
inside these branes. However, a more detailed analysis
is needed in this context.

We emphasize that BPS configuration obtained above are
different from the ones  obtained in \cite{hashi} in 
the context of noncommutative gauge 
theory. First of all, the string networks of \cite{hashi} 
preserve $1/4$ \cite{bergman}
of the $D3$ brane supersymmetry, with an interpretation 
as a dyon in these theories, whereas in our case we have 
$1/4$ of the full type II supersymmetry. Also, our string 
network lies completely inside the
$D3$ brane, compared to the ones representing dyons which connect
different branes. It will certainly be interesting to 
examine our string network configurations from the point of view of  
noncommutative worldvolume theory, appearing in this context.

These results can also be generalized to other bound states of 
fundamental and $D$-objects discussed in the literature \cite{lu,lu1} 
and will give rise to lower supersymemtries than the known ones. 
For example, one can generalize the construction of the $D1-D5$
system\cite{malda-russo} to string networks ``dissolved'' in 
$(p, q)$-webs of 5-branes. It will then be of interest to analyze 
implications of these results to black hole physics.

{\bf Acknowledgement: } We are grateful to  A. Misra and S. Mukherji 
for several useful discussions. We would also like to thank one 
of the referees of our paper for useful comments. 




\begin{thebibliography}{99}

\bibitem{witten} E. Witten,  Nucl.Phys. {\bf B460} (1996) 335,   
 [hep-th/9510135];
M. Li, Nucl.Phys. {\bf B460} (1996) 351, [hep-th/9510161]; 
M. R. Douglas, ``Cargese 1997, Strings, 
branes and dualities'' 267, [hep-th/9512077].

\bibitem{myers} J.C. Breckenridge, G. Michaud, R.C. Myers,
 Phys.Rev.{\bf D55} (1997) 6438, [hep-th/9611174].

\bibitem{malda-russo} J. M. Maldacena, J. G. Russo, JHEP {\bf 9909} (1999)
  025, [hep-th/9908134].

\bibitem{oz} M. Alishahiha, Y. Oz, M.M. Sheikh-Jabbari, JHEP {\bf
    9911} (1999) 007, [hep-th/9909215].
 
\bibitem{maldacena} J. M. Maldacena, L. Susskind, Nucl. Phys. {\bf
 B475}, (1996) 679, [hep-th/9604042];
G. Mandal, [hep-th/0002184] and {\it references therein}.

\bibitem{russo-tsey} M.B. Green, N.D. Lambert, G. Papadopoulos,
P.K. Townsend, Phys.Lett. {\bf B384} (1996) 86 [hep-th/9605146]; 
J.G. Russo, A.A. Tseytlin, 
Nucl.Phys. {\bf B490} (1997) 121, [hep-th/9611047].

\bibitem{costa} M.S. Costa, G. Papadopoulos, Nucl.Phys. {\bf B510}
(1998) 217, [hep-th/9612204]. 


\bibitem{pol} J. Polchinski, ``TASI Lectures on D-Branes'',
 [hep-th/9611050].

\bibitem{sen} A. Sen, JHEP {\bf 9803} (1998) 005, [hep-th/9711130].

\bibitem{sandip} S. Bhattacharyya, A. Kumar, S. Mukhopadhyay, 
Phys. Rev. Lett. {\bf 81} (1998) 754, [hep-th/9801141].

\bibitem{bergman} O. Bergman, Nucl. Phys. {\bf B525} (1998) 104, 
[hep-th/9712211]. 

\bibitem{hata} K. Hashimoto, H. Hata and N. Sasakura, Phys. Lett. 
{\bf B431} (1998) 303, [hep-th/9803127]; Nucl. Phys. {\bf B535} (1998) 83,
[hep-th/9804164]; T. Kawano and K. Okuyama, Phys. Lett. {\bf B432} (1998) 338,
[hep-th/9804139]; K. Li and S. Yi, Phys. Rev. {\bf D58} (1998) 066005, 
[hep-th/9804174].

\bibitem{hashi} A. Hashimoto, K. Hashimoto, JHEP {\bf 9901} (1999)
  005, [hep-th/9909202].

\bibitem{schwarz2} J. H. Schwarz, Nucl. Phys. Proc. Suppl. {\bf 55B} (1997)
1, [hep-th/9607201]; O. Aharony, J. Sonnenschein and S. Yankielowicz, 
Nucl. Phys. {\bf B474} (1996) 309 [hep-th/9603009]; 
O. Aharony and A. Hanany, Nucl. Phys. {\bf B504} (1997) 239,
[hep-th/9704170];
O. Aharony, A. Hanany and B. Kol, JHEP {\bf 9801} (1998) 002,
[hep-th/9710116].

\bibitem{dasg} K. Dasgupta and S. Mukhi, Phys. Lett. {\bf B423} (1998) 
261, [hep-th/9711094]. 

\bibitem{jabbari}  J.G. Russo, M.M. Sheikh-Jabbari, JHEP {\bf 0007},(2000) 
052, [hep-th/0006202].

\bibitem{lu2} J. X. Lu, S. Roy, JHEP {\bf 9908} (1999) 002,
  [hep-th/9904112].

\bibitem{cai} R. G. Cai, N. Ohta, Prog.Theor.Phys. {\bf 104} (2000)
 1073, [hep-th/000710]. 

\bibitem{lu1} J. X. Lu, S. Roy, JHEP {\bf 0001} (2000) 034, [hep-th/9905014]


\bibitem{schwarz} J.H. Schwarz, Phys. Lett. {\bf B360} (1995) 13, 
[hep-th/9508143].

\bibitem{kumar} A. Kumar, JHEP {\bf 0003} (2000) 010, [hep-th/0002150]. 

\bibitem{malda-ferra} J. M. Maldacena,
  S. Ferrara, Class.Quant.Grav. {\bf 15} (1998) 749, [hep-th/9706097];
A. Kumar, S. Mukhopadhyay, Int.J.Mod.Phys. {\bf A14}, 
(1999), 3235, [hep-th/9806126].

\bibitem{lu} J. X. Lu, S. Roy, Nucl.Phys. {\bf B560} (1999) 181, 
[hep-th/9904129].
 

 

     


\end{thebibliography}
\end{document}